\documentclass[fleqn,10pt]{wlscirep}
\usepackage[utf8]{inputenc}
\usepackage[T1]{fontenc}

\title{Maximally informative feature selection using Information Imbalance: Application to COVID-19 severity prediction }

\author[1]{Romina Wild}
\author[2,3]{Emanuela Sozio}
\author[1]{Riccardo G. Margiotta}
\author[2]{Fabiana Dellai}
\author[2]{Angela Acquasanta}
\author[3]{Fabio Del Ben}
\author[2,3]{Carlo Tascini}
\author[3,2]{Francesco Curcio}
\author[1,4]{Alessandro Laio*}
\affil[1]{International School for Advanced Studies (SISSA), Via Bonomea 265, Trieste, Italy}
\affil[2]{Infectious Disease Unit, Azienda Sanitaria Universitaria Friuli Centrale (ASU FC), Via Pozzuolo 330, Udine, Italy}
\affil[3]{Department of Medicine (DAME), University of Udine, Via Palladio 8, 33100, Udine, Italy}
\affil[4]{The Abdus Salam International Centre for Theoretical Physics (ICTP), Strada Costiera 11, Trieste, Italy}

\affil[*]{laio@sissa.it, Via Bonomea 265, I-34136 Trieste, ITALY, +39 040 3787426}

\begin{abstract}
Clinical databases typically include, for each patient, many heterogeneous features, for example blood exams, the  clinical history before the onset of the disease, the evolution of the symptoms,  the results of imaging exams, and many others. We here propose to exploit a recently developed statistical approach, the Information Imbalance, to compare different subsets of patient features, and automatically select the set of features which is maximally informative for a given clinical purpose, especially in minority classes. We adapt the Information Imbalance approach to work in a clinical framework, where patient features are often categorical and are generally available only for a fraction of the patients. We apply this algorithm to a data set of $\sim$ 1,300 patients treated for COVID-19  in Udine hospital before October 2021. Using this approach, we find combinations of features which, if used in combination, are maximally informative of the clinical fate and of the severity of the disease. The optimal number of features, which is determined automatically, turns out to be between 10 and 15. These features can be measured at admission. The approach can be used also if the features are available only for a fraction of the patients, does not require imputation and, importantly, is able to automatically select features with small inter-feature correlation. Clinical insights deriving from this study are also discussed. 

\end{abstract}
\begin{document}

\flushbottom
\maketitle

\thispagestyle{empty}

\section*{Introduction}
In many fields, and in particular in statistical medicine, one attempts to develop a predictor using relatively few data points (the patients), characterized by a number of features which can be  large. These features encompass demographics, vital parameters, comorbidities, medications, blood test values, radiological exams, clinical scores and more. Furthermore, they can be of any data type, \textit{e.g.} quantitative (weight, age, blood value levels), binary (presence of diabetes or other comorbidities), nominal (types of ventilation), or ordinal (sequential organ failure assessment (SOFA) score). Many of these features are typically irrelevant or redundant, namely correlated with each other, and a selection of few, relevant features is desirable. For medical professionals, having to consider too many features confuses the clinical work. 

Typically, a feature is considered relevant if it correlates with the target, for example if it discriminates between target classes. This simple concept is at the basis of most feature selection algorithms.
Feature selection methods can be broadly divided into filter, wrapper and embedded methods. Filters are simple statistics to rank the features independently of the subsequent prediction (classifier agnostic), while wrapper and embedded methods use the predictor as criterion to select feature subsets \cite{Chandrashekar2014, Yu2020}. Among the classic embedded feature selection methods, lasso-regularized regression \cite{Witten2009} provides interpretable feature selection while drawing a linear relationship between input features and target. Sparse additive models (SPAMs) \cite{Ravikumar2009} and sparse neural additive models (SNAMs) \cite{Xu2023} extend this to the non-linear case. Variable ranking filter methods tend to be easier and faster to use than other feature selection methods, but they have the drawback of being univariate methods with inability to find the optimal dimension of the feature space, namely the number of variables that are necessary to make a good prediction \cite{Chandrashekar2014, Guyon2003}.
Furthermore, many existing feature selection algorithms suffer from the inability to handle missing and noisy data \cite{Yu2020}.

In this work, we show that a recently developed approach, the Information Imbalance \cite{glielmo2021ranking}, can be used as a filter to perform feature selection in a clinical framework. 
We illustrate the procedure on a  database of  $\sim$ 1300 COVID-19 patients from Udine, including hundreds of features for each patient. These features are extremely heterogeneous, some related with the clinical history, others with the status of the patients at the admission to the hospital, other with the course of the disease, including complications, treatments and clinical outcome. Very importantly, the database is highly incomplete, as is common in clinical databases: the outcome of specific exams (say a TC scan) is typically available only for subsets of the patients, and the clinical history before the admission is often known only partially.

The Information Imbalance approach allows comparing two feature spaces, and deciding if one is more informative than the other. Feature spaces are collections of features that are used to characterize the data. For example, let's say that space A includes age, a specific comorbidity, and the value of a blood test, while space B includes the parameters measured in a TC scan and (also) age.
To estimate the Information Imbalance, one finds for each patient their nearest neighbor, which is the other patient that is most similar (closest) according to a distance estimated using the features in space A. In this study we use the Euclidean distance, but the method also allows using other distance measures. Say that for patient number 1 the most similar is patient 412, such that patient 412 has distance rank 0 with respect to patient 1 in space A. Next, one computes the Euclidean distance between patient 1 and 412 in space B, \textit{i.e.} using the features of space B, and finds the number of patients which are closer to patient 1 than patient 412. One repeats this test for all the patients and computes the average of this number. The Information Imbalance, denoted in the following $\Delta(A \rightarrow B)$, is proportional to this average. If  $\Delta(A \rightarrow B)$ is small, space A is \emph{predictive} of space B, as patients which are close in A are also close in B, and therefore the average will be taken over small distance ranks. If instead this number is large, the nearest neighbor patients in A are typically  "far" in B, which implies that space A is \emph{not informative} of space B.

A key property of this statistic is its asymmetry. At variance with correlation coefficients, the Information Imbalance can have different values when calculated from space A as reference to space B as target and vice versa. This allows capturing standard  correlations like measured by Pearson and Spearman correlation coefficients, as well as complex, asymmetric correlations, which are identified by an asymmetric Information Imbalance. 
In this work we show that Information Imbalance can be adapted to work as a filter to perform feature selection in clinical databases. 
An important advantage over other filter methods is that most statistical filter methods treat only one variable at a time. They can therefore select subsets of features which are important but redundant\cite{Kuhn2013}. More complex filter methods are designed to integrate variable relevance while minimizing variable redundancy\cite{Fleuret2004}. The approach described in this work automatically selects features which are not only relevant, but also uncorrelated. 
Indeed, the Information Imbalance can directly be computed for distances including arbitrarily many features. This is an important practical advantage with respect to other methods which are based on comparison between two  variables at a time. It also allows comparing the predictive power of subsets of features of different sizes. This, as we will see, allows finding maximally informative subsets of features along with the optimal dimension. 

\section*{Results}
This study is based on a dataset of 1308 COVID-19 patients with $\sim$ 150 features, and 33\% of missing values.
Some of these features are \emph{input} features, measurable upon admission in the hospital. These include age, gender, physical exams (\textit{e.g.} blood pressure, temperature), blood tests (\textit{e.g.} biomarkers like interleukins), arterial blood gas (\textit{e.g.} partial pressures of oxygen and $CO_2$, pH of the blood), chronic comorbidities (\textit{e.g.} diabetes) and chronic medications (\textit{e.g.} diuretics, steroids). 36\% of all input features were missing, with only 28 features being complete, and 25 features available in a quarter of the patients or less. 
The output features were 14, all binary. These features could only be measured later on during the COVID-19 infection, and are therefore the variables that a clinician might be willing to predict. These are available for all patients and include death, intubation, transfer to ICU, and 11 complications (heart attack, pulmonary embolism, arrythmia, atrial fibrillation, stroke, thrombosis, pneumothorax, pneumomediastinum, hemorrhage, delirium, and secondary infections during hospitalization). 

\subsection*{\label{sec:global}Feature selection by optimization of the Information Imbalance}
\begin{figure*}[ht!]
    \centering
    \includegraphics[width=0.7\textwidth]{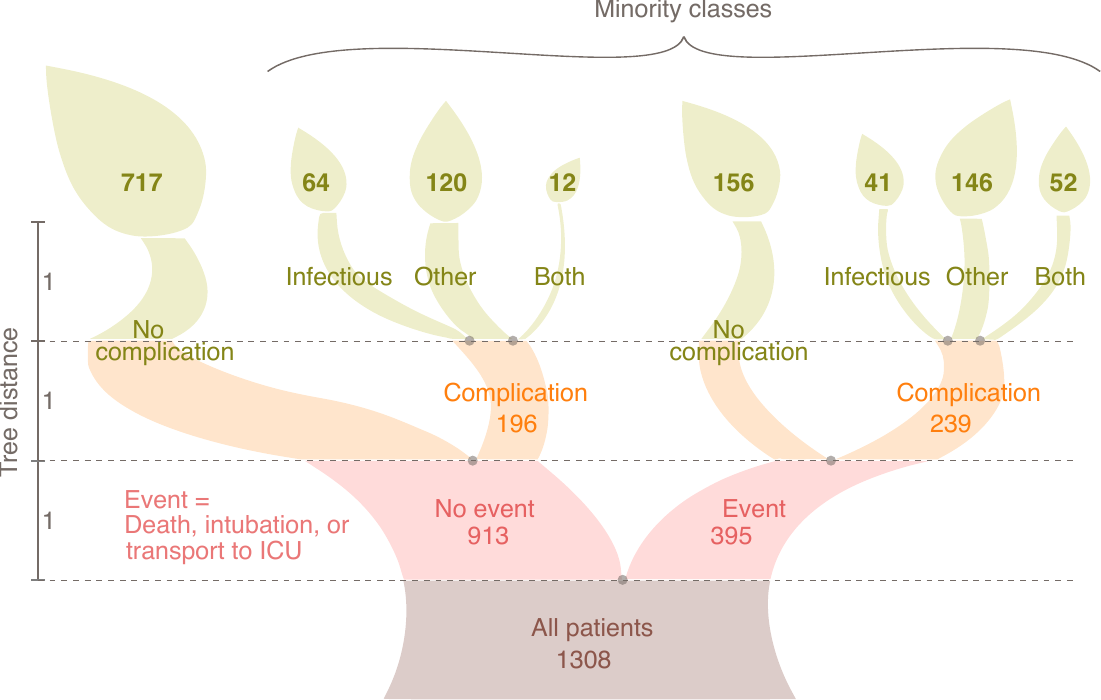}
    \caption{The severity tree splits patients into eight severity classes of different class size. These classes are depicted here as the leaves. The gray numbers on the left indicate tree distances between nodes. Distances between patients on the tree are measured one-way, such that the distance between patient A in the first big leaf (no event, no complication) to patient B in the third class (no event, other complication, 120 patients) is 2. This number is found by starting from patient A and reaching patient B: We start in leaf 1, take two steps in tree distance down to the level between orange and red, and then take two steps up to patient B's class. Counting this trajectory only downwards, the distance between the patients is 2.}
    \label{fig:tree}
\end{figure*}

We first consider a classical problem in feature selection, finding a small subset of the 138 input features which is maximally predictive with respect to the output features. In the case of the database analyzed in this work, the output features are 14. These features are all binary ("yes" or "no") and are quite heterogeneous in nature. In accordance with the medical insight of the clinicians co-authoring this study, we organized the output features in a "severity tree" (Fig. \ref{fig:tree} and Methods).

In short, death, intubation and transfer to ICU were used to split patients into two classes, one for which at least one of such events has occurred (the  patients whose course has been more severe), the other in which no event has occurred. The other output features are associated to infectious and non-infectious complications, which are important to decide a clinical strategy. We used the features associated to complications to split the two main classes into four subclasses. 

This leads to a  classification of patients into eight severity classes. The distance between two patients is then estimated by counting the number of links separating their leaves in the severity tree, divided by two (one directional). For example, the distance between a patient in class one (717 patients, no-event-no-complication) and a patient in class three (120 patients without event but with other complications) is two, while the distance between a patient in class one and in class five (156 patients with event but without complication) is three.
This tree distance is the target of the feature selection, however the nominal value of the distances is unimportant and only their relative order matters, since the Information Imbalance method uses distance ranks, \textit{i.e.} the closest neighbor in distance is assigned rank $0$, the second closest rank $1$, \textit{etc}. The feature selection algorithm presented in this work works as follows. We try to identify a distance A, built as the Euclidean distance using a combination of several input features, whose distance ranks are maximally informative with respect to distance ranks B measured on the severity tree. Degeneracies in input features were treated by addition of small random numbers (\hyperref[meth:delw]{Methods}).
The Information Imbalance between A and B is used as a feature selection filter to discriminate between different choices of A (namely of input features) and select the best one. 
The classes, "leaves" of the severity tree, are not populated uniformly: class 1 has 717 patients, and the smallest class has only 12 patients. A direct application of classic Information Imbalance in class imbalanced data sets biases class prediction heavily towards the majority class and leads to low class predictive accuracy in the minority classes, like many naive variable selection schemes in classification tasks \cite{Blagus2010}. Therefore, Information Imbalance \cite{glielmo2021ranking} has been modified by introducing class weights, aimed at compensating the occurrences of the different severity classes in the data set (Methods). We denote this modified, Weighted Information Imbalance by $\Delta_w$.

\begin{figure*}[ht!]
    \centering
    \includegraphics[width=0.99\textwidth]{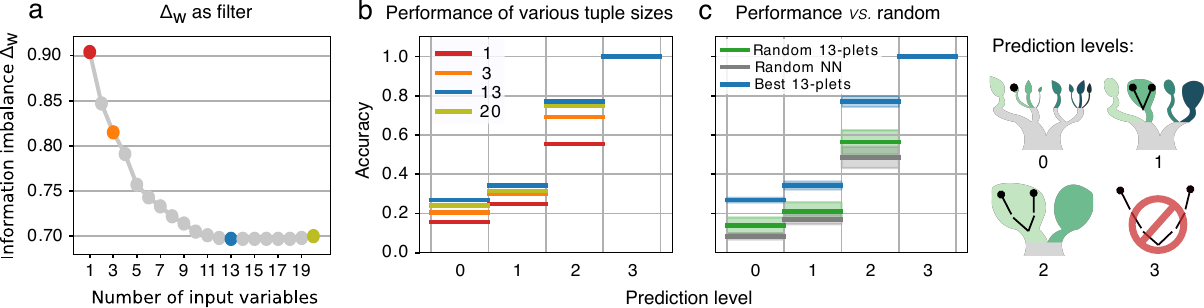}
    \caption{\textbf{a}: The optimal (lowest) Information Imbalances $\Delta_w$ as a function of the number of input features. \textbf{b}: Accuracies of 10-NN predictions at given prediction levels for tuple sizes as marked in \textbf{a} (the line shows the average over the 10 best $\Delta_w$ results for each tuple size, see main text). The accuracy corresponds to the fraction of patients predicted correctly at a given prediction level. The prediction level corresponds to the maximum tolerated distance of the true \textit{vs.} the predicted class on the severity tree, as depicted on the right. Together, the lines of one color can be considered the CDF of the fraction of patients predicted correctly. \textbf{c}: Accuracies of the top 13-plets, benchmarked \textit{vs.} randomly drawn 13-plets (green), and random nearest neighbors assignments (grey). The average over the 10 $\Delta_w$ results is shown as bold line and the standard deviation as shade. The predictions were generated by LOO. The graphic on the right explains prediction levels: $0$ denotes the fraction of patients for whom the correct class was predicted; $1$ indicates the fraction for whom the occurrence of event and complication (yes or no) was predicted correctly; $2$ means the patients predicted on the correct tree side (event \textit{vs.} no event). The remaining fraction of patients was predicted wrongly (distance $= 3$).}
    \label{fig:results1}
\end{figure*}

To identify the best combination of input features, we find the combination of $n$ input variables minimizing $\Delta_w$ , which are present in at least $100$ patients (\hyperref[meth:delw]{Methods}). For small $n$, the search can be performed exhaustively by testing all the possible combination of variables. For large $n$, the number of possible combinations grows factorially. We use the deterministic beam search algorithm (Methods) which allows finding the best combination of variables for arbitrary $n$ with great confidence. In  Fig. \ref{fig:results1}a, we plot the optimal value of $\Delta_w$ as a function of $n$.
This value decreases up to $n\simeq 13$, then starts growing slowly, indicating that adding more variables \emph{reduces} the information. This can happen when the new variables only add noise, and no independent information. This analysis indicates that the most informative combination of features includes 13 variables (blue dot in Fig. \ref{fig:results1}a). The  n-plets of features of different sizes in Fig. \ref{fig:results1}a are listed in Suppl. Inf. Section 1.

The feature space optimizing $\Delta_w$ is spanned by these features:
    \emph{ Pathologies}: Liver disease (hepatopathy);
     \emph{Chronic therapies}: Steroid therapy, potassium sparing diuretics; 
     \emph{Blood exams}: Alanin aminotransferase (GPT), antithrombin III (AT3), interleukin 10 (IL-10), troponin (TROP), antinuclear antibodies (ANA), antineutrophil cytoplasmatic antibodies (ANCA), anti smooth muscle antibodies (ASMA), Lymphocytes (FLLINF), percentage of eosinophils (FLEOS\%), percentage of neutrophils (FLNEU\%).

This combination of 13 features suggests on the one hand a systemic inflammation and autoimmunity, signaled by neutrophils and autoantibodies, and on the other hand an immune paralysis and anti-inflammatory effort (\textit{i.e.} steroid therapy, IL-10).
Furthermore, it has already been suggested that the up-regulation of inflammatory markers can lead to the progression of the disease to the severe form and eventually cause liver damage in these patients \cite{Amiri_2020}, which suggests  why hepathopathy might be an important feature. 

The information provided by these input features on the severity of the course of the disease is assessed by using these variables to predict the class of each patient. We first use a prior-corrected k-NN classifier, in which the class of a patient is assumed to be the same of their 10 nearest neighbors according to the input features (see \hyperref[meth:kNN]{Methods} for details). This predictor has no variational parameters, and lacks therefore tunability, but allows assessing directly the consistency between the neighborhood of the patients induced by our optimization procedure. Later in this study, we also use a support vector classifier (SVC) to compare the results to the k-NN classification.
In Fig. \ref{fig:results1}b, we plot the accuracies of the prediction at different prediction levels, and for a different number of features ranging from 1 to 20. The accuracy corresponds to the fraction of patients predicted correctly at a given prediction level, defined by the maximum tolerated distance of the true \textit{vs.} the predicted class on the severity tree. Level $0$ is the fraction of patients predicted completely correctly, level $1$ is the correct prediction of having an event and complication, and level $2$ is the fraction for whom only the event was predicted correctly (correct tree side). The remaining percentage belongs to patients misclassified according to having an event or not (level $3$). The accuracies are estimated by a leave-one-out (LOO) validation procedure. The performance increases with the number of features, as seen in the height of the CDF curves. This effect staggers with bigger tuple sizes and levels out.

Our approach automatically also leads to a selection of features which are practically uncorrelated with very small pairwise Pearson correlation coefficients (Supp. Inf. Section 5.1). The mean of the pairwise correlations of numerical features in the best 13-plet is $\overline{r} = 0.02$, and the pairwise Information Imbalances mean is $\overline{\Delta} = 0.96$. This happens "for free" since adding a feature which can be predicted by other already selected features does not significantly improve $\Delta_w $ (see Supp. Inf. Figure S4). 

The Weighted Information Imbalance of the top-ranking $\Delta_w$-optimized tuples of the same size is very similar. \textit{E.g.} the best 13-plet has $\Delta_w = 0.69$, while the tenth best 13-plet has $\Delta_w = 0.70$ and differs by only 3 features from the best. Therefore, in Fig. \ref{fig:results1}b and c the plotted lines are averages over the top ten results.

To put these results in context, we need to compare them to a baseline. We first performed  a comparison with the predictive performance of randomly selected tuples. 
Secondly, we performed a comparison with a prediction performed by assigning each patient a "nearest neighbor" at random. To be comparable to our $\Delta_w$-optimized tuples, both comparisons use averages over the ten best performing random 13-plets (Fig. \ref{fig:results1}c with shaded standard deviation). The prediction accuracy is much higher in the $\Delta_w$-optimized 13-plets than when using random variables. With the $\Delta_w$-optimized 13-plets the exact target class predictions are about 27\% of cases ($distance = 0$), while predicting the correct side of the severity tree (event \textit{vs.} no event, $distance = 2$) has a quota of around 77\%. In comparison, the completely random result distinguishes event \textit{vs.} no event with under 50\% accuracy and the random 13-plets from our data set predict the same with an accuracy of under 60\%.

\begin{figure*}[ht!]
    \centering
    \includegraphics[width=0.99\textwidth]{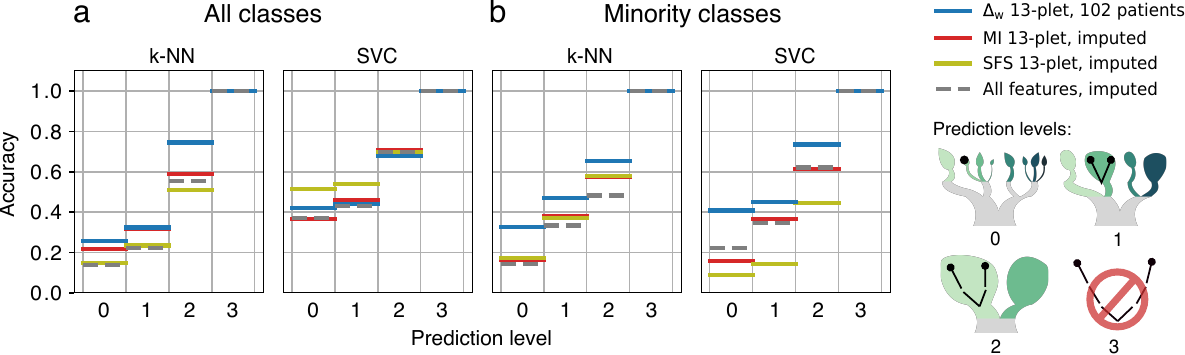}
    \caption{ Accuracy of prediction using the top $\Delta_w$-optimized 13-plet, \textit{vs.} using the 13 features with the highest mutual information (MI, filter) score on the imputed data set (red) and the 13-plet selected by forward sequential feature selection (SFS, wrapper) on the imputed data set (olive). The dashed grey line shows the prediction using the complete, imputed dataframe without prior feature selection. LOO was used for generating the predictions. Accuracy corresponds to the fraction of patients predicted correctly at a given prediction level. The prediction level corresponds to the maximum tolerated distance of the true \textit{vs.} the predicted class on the severity tree, as depicted on the right. \textbf{a}: Accuracies, using k-NN (10-NN) and SVC predictors, respectively. \textbf{b}: The same as \textbf{a}, but considering the predictions of the seven minority classes only (excluding no-event-no-complication). The SFS 13-plets for k-NN and SVC prediction are two distinct ones, using k-NN and SVC predictors accordingly in the SFS-construction of the 13-plets.}
    \label{fig:resultsSVC}
\end{figure*}

We then compared our results to standard feature selection methods, namely to selection by a mutual information (MI) score and to forward sequential feature selection (SFS). MI is a filter method which ranks the individual features against the output classes, while SFS uses the predictor method (here: k-NN and SVC) in a greedy approach to search the best n-plets, and is hence a wrapper method (see Methods). Using the top ranking 13 features selected by MI, the 13-plet selected by SFS (different for k-NN and SVC), the 13-plet selected by our approach, and all features, we perform a prediction with these four sets of features using two different approaches, the k-NN predictor with k=10 and a SVC predictor with balanced class weights. Unlike Information Imbalance, these two feature selection methods do not provide a way to select the optimal tuple size at the feature selection stage. Therefore, the predictive performances of optimal tuples of several sizes selected with MI and SFS was compared (Supp. Inf. Figure S3). Based on these results, the 13-plets were chosen, since in all three feature selection models this size is optimal or nearly optimal. The results are presented in Fig. \ref{fig:resultsSVC}a. Using the k-NN predictor the accuracy of the prediction is significantly higher if one uses our approach. Using SVC, SFS performs better than the other feature selection methods, especially at a prediction level 0 (Fig. \ref{fig:resultsSVC}a SVC). However, doing the prediction only for the seven minority classes, \textit{i.e.} excluding the no-event-no-complication class (Fig. \ref{fig:resultsSVC}b), the feature tuple obtained with our approach consistently outperforms all other tuples in both, k-NN and SVC predictions. The SFS tuple has especially low accuracy for minority classes in the SVC prediction, even though balancing class weights were employed in the tuple generation and prediction. We recall that in imbalanced multiclass prediction the problem is twofold: prediction in imbalanced datasets tends to favor the majority class, and on top of this, prediction is more complex than in binary prediction, because there can be several majority and minority classes with various relationships towards each other \cite{Tanha2020, Li2020}. Furthermore, the error introduced in the imputation could effect the minority classes more, as previously elaborated for standard imputation methods \cite{Awan2021}. 

Their reliance on imputation is an Achilles' heel of both, NMI and SFS, because most standard implementations, such as scikit-learn \cite{scikit-learn}, need data sets without missing values. In our dataset, slightly more minority class data had to be imputed than majority class data (38\% \textit{vs.} 35\%) and the different feature selection methods selected tuples with 49\% (NMI), 19\% (SFS with 10-NN) and 32\% (SFS with SVC) imputed data. Weighted Information Imbalance, as introduced in this work, uses a constrained, data-set-reductive approach (see \hyperref[meth:delw]{Methods}) and hence does not impute features. It is a non-parametric algorithm employing class-balancing weights which can intrinsically handle multiclass groundtruths. 

The prediction accuracies were also compared with two regularized classifiers, L$_1$ (lasso) regularized logistic regression and regularized sparse SVC (supplimenary Fig. S2). In terms of general predictivity of all classes these methods do well: sparse SVC even reaches over 50\% at the class level. This effect, however, does not translate to the minority classes, where this same classifier only reaches 19\% correct predictions, while Information Imbalance selection and subsequent SVC reach 41\%.

All in all, given the importance of minority classes in the clinical setting, we conclude that the Information Imbalance filter method finds a superior feature subspace for severity prediction in the present COVID-19 database than the two other feature selection methods considered here, and also then state-of-the art regularized classifiers, considering the performance in minority classes.

\subsection*{Identifying important but rarely available features}
The 13 features which have been identified as optimal as described above are simultaneously available for only 102 patients. Indeed, approximately one third of the data are missing in the data set. Moreover, missing values are not evenly distributed among the features: some "cheap" exams are performed routinely for all  the patients, others are performed only for a small fraction of the patients. As a consequence, optimal features might not be available for a generic patient. 

We repeated the optimization procedure by attempting to select, independently for each patient, the best tuple among the features which are available for them. We beam-searched over the possible feature tuples for the respective patient, using the subset of the patients which had no missing values in these features (see Supp. Inf. Section 2). The patient-specific optimal tuple was subsequently used in a k-NN prediction, using the other patients with full information as training set. 
The average predictive performance of the features selected in this manner is still very significant but is reduced as compared to what observed for the optimal 13-plets (Supp. Inf. Figure S1). The prediction of the correct side of the tree (event \textit{vs.} no event) is reduced by about 7\% to roughly 70\% and the correct class prediction drops from 27\% to about 23\%. This result is not surprising: the variables which turn out to be most informative happen to be simultaneously available only for a relatively small fraction of patients.

This result pushed us to develop a quality measure for the input features which takes into account the fact that  $\Delta_w$-optimized n-plets contain features that are "good" in two ways: firstly, they are intrinsically important, and secondly, available together in the same patients. 
We  estimate, for each feature $f$, the  number of patients $NP_f$ for which $f$  is included in the most informative  $\Delta_w$-optimized tuple (which is patient-specific). The usage $U_f$ of feature $f$ is then estimated by the ratio between $NP_f$  and the number of patients for which $f$ is available. If $U_f$ is close to one, then the feature has been chosen for each patient where it was available, and has a high intrinsic importance for the Information Imbalance towards the severity tree target classes. 

\begin{figure*}[ht!]
    \centering
    \includegraphics[width=0.88\textwidth]{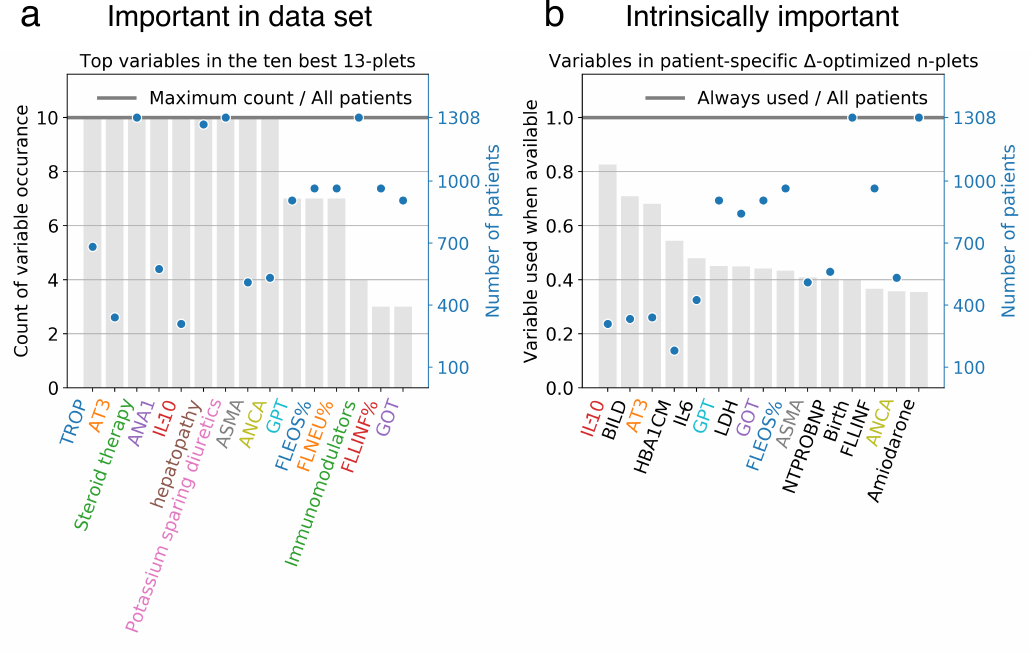}
    \caption{\textbf{a}: Statistics of the features present in the best 10 13-plets. The grey bar indicates the fraction of 13-plets in which the variable is present.  The blue dots indicate in how many patients the feature is available (scale defined on the right y-axis). \textbf{b}: Intrinsically important features estimated by the  "usage when available" statistic $U_f$. The gray bars indicates the value of $U_f$, which is large when a variable is always used when present. The blue dots are the same as in panel A.}
    \label{fig:best_f}
\end{figure*}

Fig. \ref{fig:best_f}a shows the features which are used in the 10 most predictive 13-plets (predictive performance of these in Fig. \ref{fig:results1}c) which, we recall, are present at the same time in $\sim$ 100 patients. 
9 of the 13 features are used in all the ten best models. The blue dots indicate the fraction of patients for which the variable is available. For example, information about steroid therapy, which is used in all the 10 best models, is available for all the patients. AT3, also used in all the best models, is available for approximately 350 patients. 
Fig. \ref{fig:best_f}b on the other hand shows the value of the "usage when available " statistic $U_f$. 
Some of those variables are present in both sets of Fig. \ref{fig:best_f}, namely the cytokine IL-10 (interleukin 10), the anticoagulant protein AT3 (antithrombin III), the autoantibodies ANCA and ASMA (antineutrophil cytoplasmic antibodies and anti-smooth muscle antibody), the liver enzymes GPT and GOT (alanin aminotransferase and aspartate aminotransferase), and the percentage of eosinophils, FLEOS\%.
The $U_f$ statistic can also find intrinsically good predictors which are underrepresented due to missing values,  and as such do not appear in the most used features of the $\Delta_w$-optimized tuples. For the COVID-19 severity prediction some of these are direct bilirubin (BILD), the diabetes indicator glycated hemoglobin (HBA1CM), interleukin 6 (IL-6), and the enzyme lactate dehydrogenase (LDH), which indeed have a high value of $U_f$ (panel b) but do not enter in the best model (panel a). IL-6 is a pro-inflammatory cytokine and has previously been linked with COVID-19 severity \cite{Fabris2022}. Abnormal bilirubin levels indicate sepsis, and the severity of patients could be linked to the fact that they have developed sepsis, and therefore a condition of a dysregulated systemic response. Glycated hemoglobin is linked to a condition of decompensated diabetes, which predisposes to infections. Diabetes is known to predispose severe COVID-19 infections \cite{CDC_2022}.

\section*{Discussion}

In this work, we show that the Information Imbalance \cite{glielmo2021ranking} approach can be used to perform feature selection in a clinical database. We illustrated our approach by the analysis of a database of 1300 COVID-19 patients from Udine hospital with $\sim$ 150 features for each patient and one third of missing data.
In order to deal with unbalanced classes in a clinical setting we had to include weights in the Information Imbalance definition. We find that the optimal feature tuples selected by our approach  perform better in the k-NN prediction of COVID severity classes than two other standard feature selection methods, mutual information (a filter) and sequential feature selection (a wrapper), implemented in a standard statistical analysis package\cite{scikit-learn} (see Fig. \ref{fig:resultsSVC}). Regulated classifiers which do not include a prior feature selection step can reach slightly higher accuracies for the prediction of all classes (Supp. Inf. Figure S2). This effect however does not translate to minority classes, which in our database correspond to patients developing very severe symptoms.  Information Imbalance feature selection followed by SVC classification outperforms all other methods in identifying those patients.

The classification task, as applied in our study, serves to validate the utility of the $\Delta_w$ selected features in distinguishing between severity classes of COVID-19, rather than introducing a complete novel classification model. A future direction of research could extend this work into a framework including techniques for quantifying of the uncertainty of prediction, but is beyond the scope of the current work and would overshadow the core contribution of our novel feature selection method.

The optimal variables of our data set include cytokines (such as IL-10), autoantibodies (ANA, ASMA, ANCA), and therapies that reduce the immune response (steroid therapy and immunomodulators as chronic home therapy). These findings are of medical interest because the pathogenic mechanisms that drive COVID-19 clinical deterioration can likely be contributed to systemic inflammation, disordered coagulation (AT3), and immune dysfunction.
The cytokine storm that characterizes the unfavorable outcome of patients comprises classical markers of systemic inflammation such as IL-6, which is now largely disposable as a single diagnostic test also at urgent request in the majority of hospitals around the world.  IL-6 increases during COVID-19 illness decline as patients recover, correlating with the severity of the disease course \cite{Fabris2022}. When IL-6 levels are already very high, the focus can be shifted to the degree of immunoparalysis and anti-inflammatory effort (IL-10) \cite{Fabris2022}.
COVID-19 is known to alter the coagulation state and, in severe cases, lead to hypercoagulation, which is causally involved in negative patient outcomes. AT3 levels decrease in inflammatory conditions and AT seems to possess anti-viral properties \cite{Schlommer2021}.
The role of transaminases (GPT and GOT), and history of liver disease are also interesting. The hepatic consequences of an SARS-CoV-2 infection are recognized as an important component of COVID-19 and this aspect is most clinically relevant in patients with pre-existing cirrhosis \cite{Marjot_2021}. There are several other potential contributors to abnormal liver biochemistries in COVID-19, including ischaemic hepatitis, hepatic congestion related to cardiomyopathy, and transaminase release due to the breakdown of skeletal and cardiac muscle \cite{Jothimani2020}.

As common in most real world clinical data, also in this case the accuracy does not allow to perform exact prediction of the patient fate into their severity class (prediction level $0$ in Figs. \ref{fig:results1} and \ref{fig:resultsSVC}). Exact predictions, as tested in a leave-one-out-cross-validation approach, happen in   30\% - 40\% of cases, depending on which predictor is used and which classes is considered. However, in over 70\% of the cases we were able to predict if the patient will suffer a serious event (death, transport to ICU or intubation, prediction level $2$) or not. In this sense, the hierarchical tree structure of the output space is of advantage, because even if the exact prediction is not possible, a warning along a less stratified prediction level of the output tree is nevertheless possible. 
The feature tuples derived from our approach keep the high accuracy described above also for minority class patients, as opposed to tuples generated with other feature selection techniques. 

One important advantage of the approach presented here is that it works without imputation\cite{Donders2006}, namely it does not require the preprocessing step of assigning missing values of the input features. While MI and SFS in their standard implementations need complete data sets and such require imputation, our approach finds patient-specific optimal input feature tuples in the original, incomplete data.

A limitation of the method is presented by the nature of the ground truth space. Since Information Imbalance finds input feature spaces which reproduce the neighborhood relationships observed in a ground truth space, it works best if the ground truth is either continuous, or at least has classes for which one can meaningfully identify a distance similar to our severity tree. The method is less suitable for learning a ground truth distance which can take only a few values, and would not be appropriate for binary classification tasks. Furthermore, continuous input features are more likely to be chosen in small tuples, because they can carry more information than categorical variables. The optimal 13-plet of this study includes three binary variables (hepatopathy, steroid therapy, potassium sparing diuretics: yes/no
), proving that there are several informative binary features in this COVID-database that hold information complementary to the chosen numerical features.

The threshold of $ \geq 100$ patients without missing data and Jensen-Shannon divergence of $\leq 0.06$ for Information Imbalance feature selection were a practical choices, which can influences the specific results of the optimal tuple. While this choice has proven to balance performance and representativeness effectively, different informative feature tuples could be selected for different thresholds. The optimal 13-plet should be considered one out of several possible informative feature combinations in this data set.

This work also tested the features for their intrinsic importance decoupled from their availability in the data set by employing a simple usage statistic, which we called $U_f$.
This analysis can be used to provide  recommendations to clinicians for future data collections, because it identifies potentially important features for COVID severity prediction, which are, however, not abundant enough in our data set. For example, we find that  conjugated (direct) bilirubin (BILD),  is available for less than 30\% of the patients in the current data set, yet it is selected very often, when present, for the patient-optimal prediction. A similar scenario is found for glycated hemoglobin (HBA1CM) and interleukin 6 (IL-6). The collection of these features should be emphasized in future data collection efforts.

The use of clinical features and diagnostic features as biomarkers is of great clinical interest, in order to facilitate improved triaging and earlier therapeutic decisions. 
The model presented here could help the clinicians to focus on the variables of greatest interest in order to target the allocation of resources and escalation of care.

\section*{Methods}

\subsection*{The data set}
This work is a retrospective study involving data of 1308 COVID-19 patients from Udine hospital. The data set includes patients admitted to the Infectious Disease ward of the Azienda Sanitaria Universitaria Friuli Centrale Santa Maria della Misericordia of Udine, a 1000-bed tertiary-care teaching hospital identified as a regional referral center for COVID-19 patients, from March 2020 to March 2021. Informed consent was obtained from all participants (see section \hyperref[sec:informed_consent]{Informed consent}).

For all patients the following parameters were collected: evaluation of in/exclusion criteria; socio-demographics (age, gender, race, height, weight); date and time of the onset of symptoms and of the admission to the hospital; ward of hospitalization; co-morbidities (dyslipidemia, obesity, diabetes, chronic obstructive pulmonary disease, chronic kidney injures, liver disease, hypertension, solid and hematologic neoplasms, autoimmune diseases, primary or secondary immunosuppression), including Charlson score index; findings from routine physical examination (temperature, heart rate, breathing rate, blood pressure, SPO2, neurological status); routine diagnostics performed (chest X-ray, CT scan, ultrasound, microbiological tests and blood tests performed); date and time of blood sampling initial and final diagnosis; type and focus of infection; date of discharge; date and time of ICU admission and discharge; needs for organ support and/or invasive ventilation; any serious adverse event or complication which occurred during hospitalization; therapies carried out; lab parameters from routine blood testing which were assessed at presentation (within 48 hours of admission); data from blood gas analysis, such as  PaO2/FiO2 ratio, alveolar arterial gradient, and lactate.

The features were divided into input (measurable upon hospitalization) and output (severity of outcome of COVID infection) features.
14 output features were decided upon by medical knowledge. The natural hierarchy in the severity of these features led to the creation of a "severity tree", where death, intubation and transfer to ICU were used to split patients into an "event" group, where at least one of these three events occurred, and a "no event" group. Various infectious and non-infectious complications make up the second split of the tree, as seen in Fig. \ref{fig:tree}. 
These output features were available for all patients, yielding a distance (and an identical distance rank) between 0 and 3 for each pair of patients on the tree. Since there are only eight  classes in the severity tree, the distances between all patients in this output space are degenerate, meaning that many patients have the same distances from each other. Furthermore, there is a high class imbalance, \textit{i.e.} the biggest class contains more than half of the patients.

138 input variables were selected, including age, gender, physical exams (\textit{e.g.} blood pressure, temperature), blood test (\textit{e.g.} diagnostic antibodies and interleukins), hemogas (\textit{e.g.} partial pressures of oxygen and $CO_2$, pH of the blood) values and chronic comorbidities (\textit{e.g.} diabetes) and medications (\textit{e.g.} diuretics, steroids). Potential input spaces are tuples of these input variables.
As many real world data, the input data is characterized by missing values, which leads to reduced patient numbers for certain combinations of input features. In some possible input spaces also distances between data points are degenerate, due to categorical, binary and repeated values.

\subsection*{The Information Imbalance}
The  Information Imbalance $\Delta$ is defined as follows (for more details see \cite{glielmo2021ranking}):
\begin{equation} \Delta(A \to B) \approx \frac{2}{N} \langle r^B \vert r^A = 1 \rangle.
\label{eq:II} \end{equation}

This means that $\Delta$ is a scaled mean of the distance rank of a specific neighbor of each data point in space $B$ ($r^B$), where that specific neighbor is the nearest neighbor (has distance rank 1, $r^A = 1$) of the data point in space $A$. Here, a data point is a patient. The rank is the distance position of one point to another in relation to all other points, \textit{i.e.} the nearest neighbor of a point has rank 1 to that point, the second nearest neighbor rank 2, in the original formulation. Please note that due to the tree structure in the COVID-19 output space in this paper we assign the nearest neighbor rank 0, which leads to very small numeric differences in the case of few data points. Several distance metrics could be used to calculate these ranks. Since the method is focused on the identification of a feature space which reproduces the nearest neighbors of another feature space, it is not very sensitive to the precise choice of the distance metric. While the distance between two 'far' points will likely be very different if computed, \textit{e.g.}, with the Hamming distance or with the Euclidean metric, the nearest neighbors are more preserved across metrics. This work uses the Euclidean distance. $\langle \cdot \rangle$ denotes the expectation value, the arithmetic mean. N is the number of all data points, in this case all patients. Equation \ref{eq:II} is estimated as
\begin{equation} \Delta(A \to B) = \frac{2}{N} \frac{\sum_{i=1}^N r^{B'}_i}{N},
\end{equation}
where $r^{B'}_i = r^{B}_i$ given $r^{A}_i = 1$. 
The Information Imbalance between feature space A and feature space B, $\Delta(A \rightarrow B)$, is proportional to the average of the neighbor ranks in space B, conditioned to nearest neighbors in space A, and normalized such that if A predicts space B perfectly, $\Delta(A \rightarrow B) \approx 0$, and if A has no information on B, $\Delta(A \rightarrow B) \approx 1$.

\subsection*{Class-corrected Weighted Information Imbalance}
\label{meth:delw}
For the high dimensional classification of COVID-19 severity, Information Imbalance is used in the feature selection step. 
The task is to find input variable spaces in which the nearest neighbors optimally describe the distribution of the output classes. The severity tree output space is degenerate with high class imbalance, where leaf 1 has more than half of the total patients. Using naive classifiers in class imbalanced data sets biases class prediction heavily towards the majority class, especially when feature selection is performed. Subsequently, the class predictive accuracy is low in the minority classes \cite{Blagus2010}. The original implementation of Information Imbalance was developed for continuous in- and output spaces and has the described shortcomings when applied to few, imbalanced output classes.  However, in sick patient prediction there is a high cost associated to false negatives. To find feature spaces which also predict the neighborhood of the patients in small classes, we introduce class weights $w_{i, l} = \frac{1}{leafsize_l}$, for each patient $i$ in class $l$.
Furthermore, the distances in the severity tree are highly degenerate. The structure of our tree leads to a total of four distinct distances and identical four distance ranks, where rank $0$ means that two patients are in the same class and rank 3 means the two patients are on opposite sides of the tree (event \textit{vs.} no event). Since the classes are imbalanced, the average probability to find a certain rank neighbor from the different classes is not uniform. Therefore, the normalization $a$ is built to reflect this and bring the average value of $\Delta$ to a value of $1$, when the nearest neighbor ranks are distributed randomly. The adjusted "Weighted Information Imbalance $\Delta_w$" becomes:
\begin{equation} 
\Delta_w(A \to B) \approx a \frac{\sum_{i=1}^N r^{B'}_i w_i}{\sum_{i=1}^N w_i} 
\end{equation}
If more than one nearest neighbor exists at the same distance in the output space B, the nearest neighbor rank is the mean over these M nearest neighbors of patient $i$: $r^{B'}_i = \frac{\sum_{j=1}^M r^{B'}_{i,j}}{M}$.

For this implementation the input space needs to provide clear nearest neighbor assignments, in order to find the according ranks in the output tree. To resolve the degeneracy in the input space, small random numbers are added to duplicated input values. Since this makes the estimated imbalance a stochastic variable, we repeated the optimization ten times with different random seeds, verifying that the results are robust. 
For the ten implementations, the Weighted Information Imbalances of optimal tuples with the same tuple sizes are mostly identical, up to the second digit (equal to figure \ref{fig:results1}a) with standard deviations on the order of $10^{-4}$. Also the chosen tuples for each size are largely congruent, with the best single variable always being brain natriuretic peptide (BNP) and the best 13-plets of the ten implementations being identical to the one present in this study, except in one case, where eosinophils (FLEOS) have been selected instead of lymphocytes (FLLINF). 

Variables were normalized by dividing them by their standard deviation in order to move them into a comparable value range. 

The problem of missing data was treated in a constrained, data-set-reductive manner, by which only feature tuples were considered which were present in at least 100 patients. Then $\Delta_w$ was calculated for the feature tuple in question using only these $\geq 100$ data points with the additional constraint that the base-2 Jensen-Shannon divergence of the tuple class distribution towards the full class distribution (of the 1308 patients) was $\leq 0.06$ , in order to ensure proportionate stratified sampling, \textit{i.e.} such that the share of the classes in the sub-sample is proportionate to the full sample. This subset of $\geq 100$ patients has no missing values and can hence be passed on to the downstream classifiers, kNN and SVC, without problems.

Input spaces with more than 1 variable were selected by applying $\Delta_w$ to a pool of candidate feature tuples selected by beam search with beam width 55. Beam search is a heuristic algorithm\cite{Furcy2005} which is employed because the full exhaustive search of all possible variable $k$-tuples (out of $D$ features) would lead to combinatorial explosion with complexity $\mathcal{O}(D^k)$. Like in the vanilla greedy approach, the pool of candidate feature tuples is sequentially extended by adding new features to the best scoring previous tuples. However, in beam search, not only the one best result is chosen and variables added to it, but the $n$ best results, where $n$ denotes the beam width. In this case, the 55 best results were iteratively extended. The computational complexity of beam search is $\mathcal{O}(D \cdot n \cdot k)$ \cite{Zhang2023} for beam width $n$.

\subsection*{Prior-corrected k-NN prediction of severity}
\label{meth:kNN}
After finding $\Delta$-optimized sets of features, we use an adjusted k-nearest neighbor (k-NN) prediction in a leave-one-out (LOO) approach. Comparing the predicted severity class to their actual class, we evaluate the performance of the method with cumulative distribution functions (CDFs) of the distances $d$, \textit{i.e.} consider the fractions of cases in which the class was predicted correctly ($d = 0$), or a neighboring class was predicted ($d = 1$), or the same side of the severity tree was predicted ($d = 2$).
The empirical probability distribution of classes for each patient calculated from their nearest neighbors' classes cannot be taken at face value due to the class imbalance. Effectively, the class imbalance makes it much more likely to find a majority class NN than to a minority class NN. The average global density of minority class points is smaller. For this reason we divide the empirical probabilities by the prior probabilities $P^0$ of the classes in the samples, to create a metric which measures how much bigger or smaller the local density of the various classes is around the patient, in comparison to the average global density. If this value is greater than 1 for a class, this class is more likely than average to be the class of the patient. Since several classes can have values greater than one, the prediction is based on the class with the maximum value. 

\begin{equation} P^0_{i,l} = \frac{leafsize_{i,l}}{\sum_{k=1}^{Nl} leafsize_{i,k}} \end{equation}

\begin{equation} \rho_{i,l} = \frac{P^{emp}_{i,l}}{P^0_{i,l}} \end{equation}

\subsection*{Identifying important but rarely available features by “usage when available”}
Due to the missing values in the input data set, the globally selected best variables are a function of their intrinsic goodness of prediction as well as their availability, especially together with other orthogonal features. To decouple these effects to find intrinsically valuable features, 
the optimal tuple for each patient is found using  $\Delta_w$ in a leave-one-out (LOO) procedure. 
The "usage when available" statistic $U_f$ for each feature $f$ is simply defined as the ratio of the number of times a feature is used in all patient-specific $\Delta_w$-optimized tuples $n_{f,\Delta_w}$, over the count of the availability of that feature across all patients $a_f$:

\begin{equation}  
U_f = \frac{n_{f,\Delta_w}}{a_f}
\end{equation}

\subsection*{Mutual information and sequential feature selection}
Two other standard feature selection methods are compared to Information Imbalance. A frequently used filter is the estimated mutual information (MI) between each feature and the output classification \cite{Kraskov2004}, and a straight forward wrapper method is sequential feature selection (SFS). For both the implementation in scikit-learn \cite{scikit-learn} is employed and for both methods the data set has to be complete. Thus missing values were filled in by imputation (scikit-learn KNNImputer with 10 NN, uniform weights and Euclidean distance). The MI between each feature and the output classes were calculated with the scikit-learn class mutual\_info\_classif (3-NN). Forward SFS was calculated using the wrapper SequentialFeatureSelector (5-fold, 10-NN) and SVC (balanced class weights) respectively). The prior-corrected k-NN predictor was used for classification in a LOO approach as explained above, and compared with a support vector classification (SVC) LOO prediction as implemented in scikit-learn (sklearn.svm.SVC), using default settings and balanced class weights.

\section*{Informed consent}
\label{sec:informed_consent}
From March 2020 to March 2021, clinical data were collected in the retrospective registry “MAnagement coroNavirus Disease In hospital registry” – MANDI registry –  authorized by the regional ethics committee Comitato Etico Unico Regionale del Friuli Venezia Giuli (CEUR FVG) (decree n. 957, 10/09/2021). Inclusion criteria included age of at least 18 years and given informed consent to use the de-identified clinical data. All research was performed in accordance with the Declaration of Helsinki.

\section*{Data availability}
The COVID-19 patient data extracted from the MANDI registry, used for these analyses, as well as the code of the Weighted Information Imbalance optimization, is available at Open Science Framework: \url{https://osf.io/b4ktd/?view_only=2a71edf707bc4df78cb4d3b84c889461}

\bibliography{bib}

\section*{Author contributions statement}
R.W., E.S., R.G.M. and A.L.  performed  research. R.W., E.S., R.G.M. and A.L. wrote the manuscript and prepared the figures. A.L., R.W., E.S., R.G.M., F.D.B, C.T. and F.C. designed the research. E.S., F.D.B., C.T., F.C., F.D. and A.A. collected and curated the data.  

\section*{Additional information}

\textbf{Competing interests:} The authors declare no competing interests.

\end{document}


\maketitle

\section{Optimal feature sets of different sizes}
The globally best n-plets  of features as a function of n  (until n=13), corresponding to Fig. 2a of the main paper: 
\begin{enumerate}
\item BNP
\item AT3, IP10
\item TC, A-a gradient, IP10
\item FLNEU\%, AT3, IP10, ASMA
\item FLNEU\%, AT3, IP10, ACARG, ASMA
\item FLLINF\%, AT3, IL-10, TROP, ANCA1, ASMA
\item GOT, FLNEU\%, AT3, IL-10, TROP, ANCA1, ASMA
\item Steroid therapy, FLNEU\%, GPT, AT3, IL-10, TROP, ANCA1, ASMA
\item Hepatopathy, FLLINF\%, GPT, AT3, IL-10, TROP, ANA1, ANCA1, ASMA
\item Hepatopathy, FLLINF\%, GPT, AT3, IL-10, ACARG, TROP, ANA1, ANCA1, ASMA
\item Hepatopathy, steroid therapy, potassium sparing diuretics, FLLINF\%, GOT, AT3, IL-10, TROP, ANA1, ANCA1, ASMA
\item Hepatopathy, steroid therapy, potassium sparing diuretics, FLEOS\%, FLNEU\%, GPT, AT3, IL-10, TROP, ANA1, ANCA1, ASMA
\item Hepatopathy, steroid therapy, potassium sparing diuretics, FLLINF, FLEOS, FLNEU\%, GPT, AT3, IL-10, TROP, ANA1, ANCA1, ASMA
\end{enumerate}

\section{Predictive power for patients without the optimal input tuples}

\begin{figure}[ht!]
    \centering
    \includegraphics[width=0.66\textwidth]{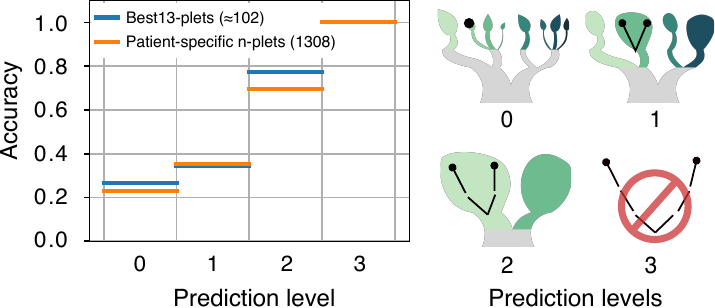}
    \caption{Accuracies of 10-NN predictions at given prediction levels by the optimal 13-plets (averaged over the ten best), which are available in roughly 102 patients, \textit{vs.} accuracies using for each patient in the database (1308 patients) their optimal input feature tuple. The accuracy corresponds to the fraction of patients predicted correctly at a given prediction level. The prediction level corresponds to the maximum tolerated distance of the true \textit{vs.} the predicted class on the severity tree, as depicted on the right. Together, the lines of one color can be considered the CDF of the fraction of patients predicted correctly.}
    \label{fig:s_non_opt}
\end{figure}

The optimal 13-plet from our study is only available for about 100 patients. Our method, however, can find for each patient their patient-specific best n-plet, with a slight loss of predictive accuracy when averaged over all patients (Fig. \ref{fig:s_non_opt}). This makes sense since 102 patients have the features of the best 13-plet, while the rest the patients do not have complete data for these features. Hence, their respective $\Delta_w$-optimized n-tuple has higher (worse) Information Imbalance than the optimal 13-plet, which influences the prediction accuracy. 

The patient-specific optimal n-plets were found in a leave-one-out (LOO) approach by considering all features that were present in the respective patient, then beam-searching over these starting from the 1-plets. For each of these feature tuples the Weighted Information Imbalance is calculated using all the patients who have full information in these features, and the search is stopped when the Information Imbalance flattens or starts increasing. In this way, the patient-specific optimal tuple is found, and along with it the optimal dimensionality. Then we performed a 10-NN prediction of severity for each patient, using their optimal n-plet of features in a LOO cross validation, where we use all other patients who share the same features as training set. As default, the algorithm only considers possible feature tuples which are available in at least $100$ patients and have a base-2 Jensen-Shannon divergence of $\leq 0.06$, in order to be representative of all classes in the full set.

\section{Accuracy of prediction compared to regularized classifiers}
\begin{figure}[ht!]
    \centering
    \includegraphics[width=0.99\textwidth]{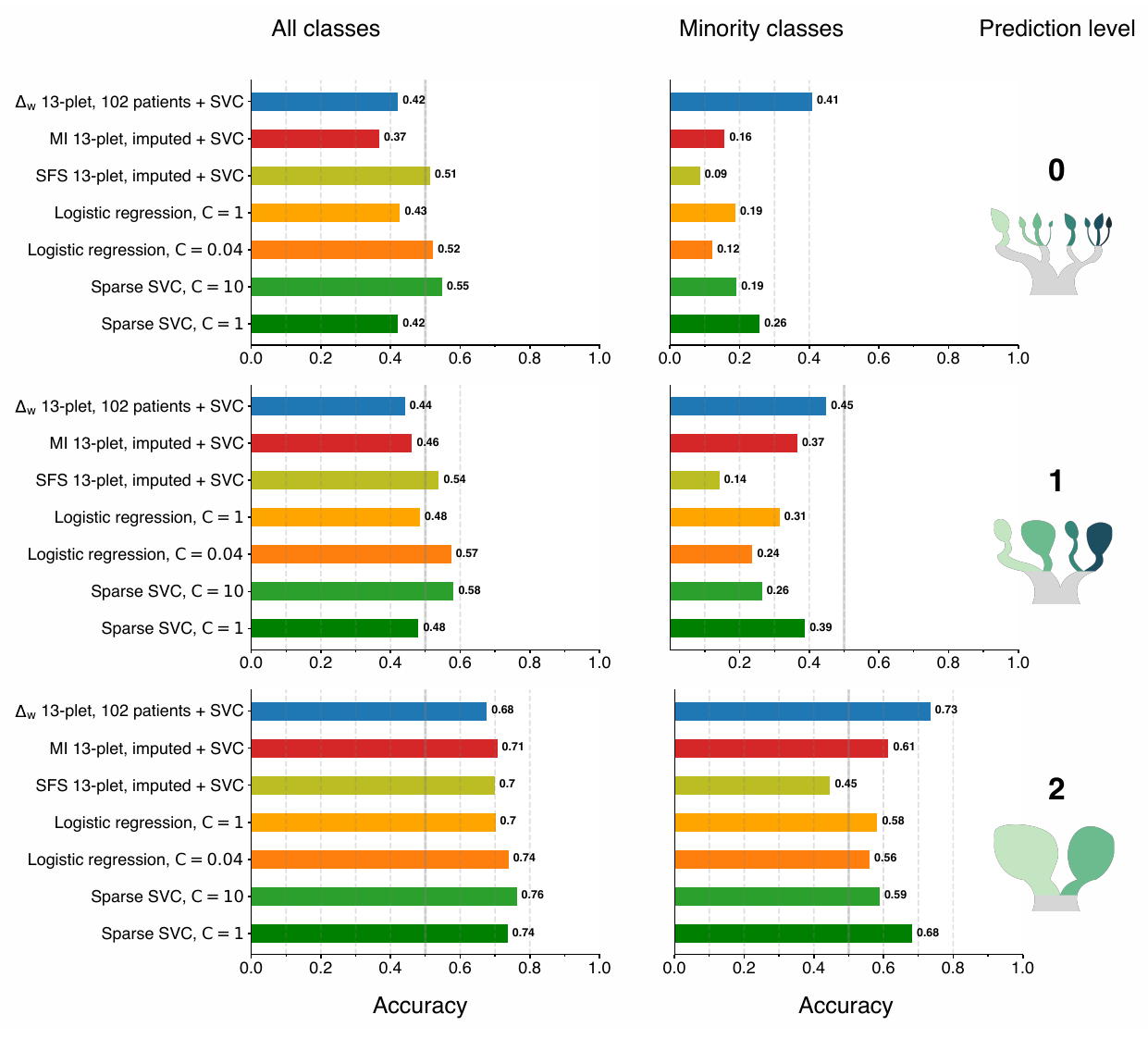}
    \caption{Accuraciess of prediction of several models, corresponding to Fig. 3 of the main paper (SVC only), and compared with two classifiers, logistic lasso regression and sparse SVC. The accuracy corresponds to the fraction of patients predicted correctly at a given prediction level. The prediction level corresponds to the maximum tolerated distance of the true \textit{vs.} the predicted class on the severity tree, as depicted on the right. 0: exact class predicted; 1: event and complication predicted; 2: event predicted.}
    \label{fig:s_classif}
\end{figure}

Information Imbalance is not a classification method. However, in the application described in this work the ground truth metric is defined on a severity tree. Its leaves can be considered as categories (classes) which can be used as a target for a classification method. The approach in the main paper introduces class-corrected Weighted Information Imbalance as a novel filter method to do feature selection, which is followed by classification. Here, a comparison with regularized classifiers, which do not employ explicit prior feature selection, is presented. The sklearn \cite{scikit-learn} implementations of 
\begin{enumerate}
    \item L$_1$ (lasso) regularized logistic regression classification \\sklearn.linear\_model.LogisticRegression with \\penalty='l1', C=1 or 0.04, max\_iter=200, class\_weight='balanced', solver='liblinear', tol=0.01
    \item Regularized sparse SVC \\sklearn.svm.SVC with C=10 or 1, kernel='rbf', gamma='auto', class\_weight='balanced'
\end{enumerate}
were tested.
The results are presented in Fig. \ref{fig:s_classif} and build on Fig. 3 of the main paper. Indeed, in terms of general prediction accuracy these regularized classification methods perform at least as good, sometimes better, as the filter models. When considering minority classes only (leaving out the biggest class of patients who did not have a complication or event), then Information Imbalance followed by SVC classification outperforms all other models, especially for the most exact prediction level (0). For all prediction levels, modestly regularized ($C=1$) sparse SVC yields the second best results.

\section{The performance of other approaches using feature tuples of various sizes}
\begin{figure}[ht!]
    \centering
    \includegraphics[width=0.9\textwidth]{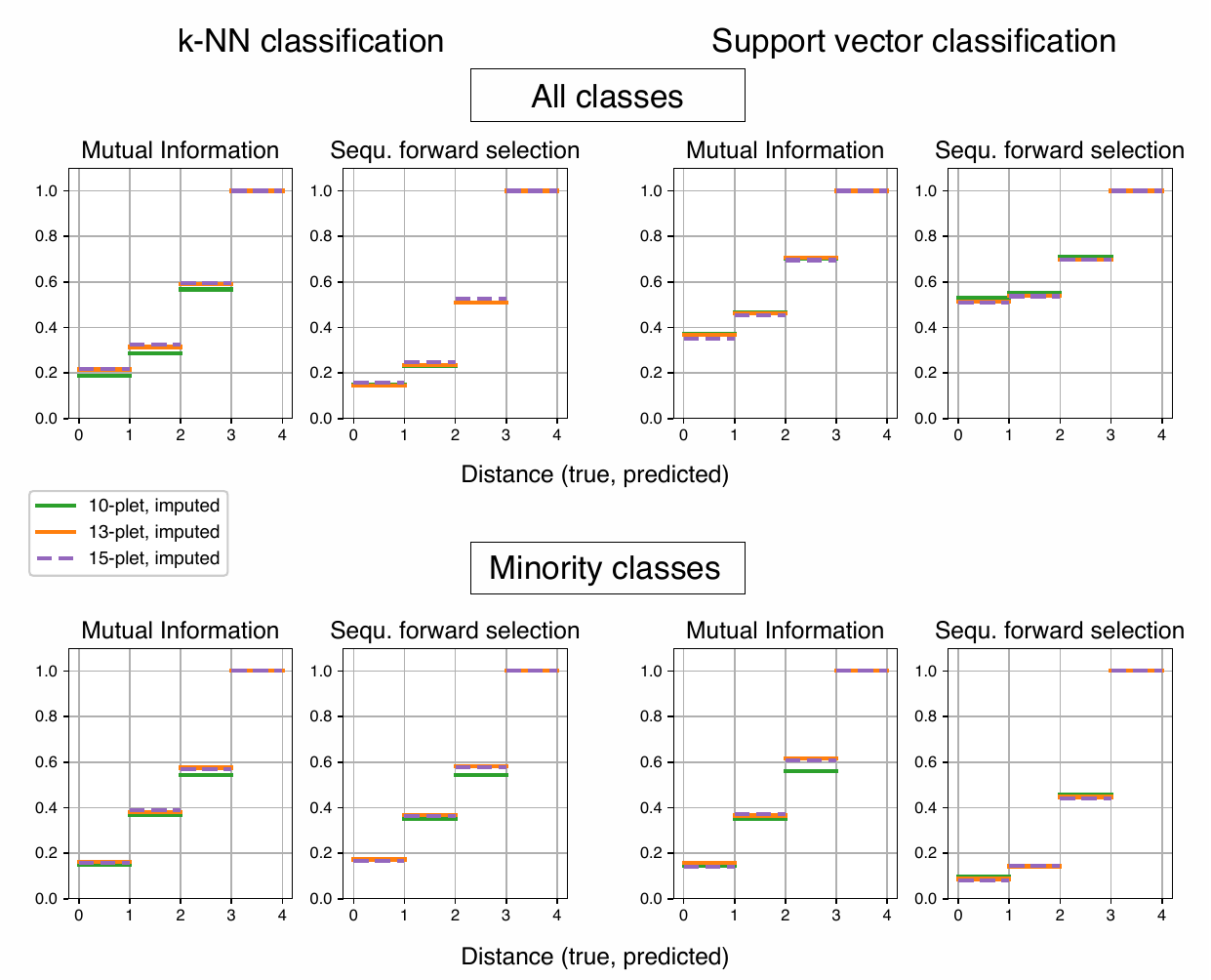}
    \caption{Accuraciess of k-NN and SVC predictions using optimized tuples of different sizes. The tuples were optimized by feature selection with Mutual Information (MI) and sequential forward selection (SFS). The accuracy is described in Fig.\ref{fig:s_classif}.}
    \label{fig:tuple_sizes_mi_sfs}
\end{figure}

The algorithm described in this work provides an automatic way to determine the optimal size of the predictive tuple at the feature selection stage - an advantage that other filter methods do not have. In the main paper we compare WII selected 13-plets with 13-plets selected with two other methods, Mutual Information (MI) and sequential forward selection (SFS). Since these two methods do not provide a determination of optimal tuple size at the feature selection stage, the comparison can be performed only at the prediction accuracy stage. 
The results of this analysis are shown in Fig. \ref{fig:tuple_sizes_mi_sfs}. 
The performance of these approaches is almost independent on the tuple size, for sizes between 10 and 15.

\section{Correlation and Information Imbalance between input features}

\subsection{Information Imbalance and correlation within $\Delta_w$-optimized n-plets}
\begin{figure}[ht!]
    \centering
    \includegraphics[width=0.4\textwidth]{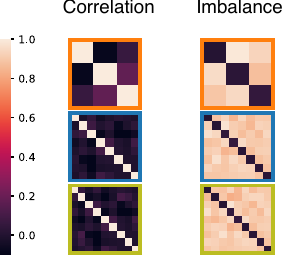}
    \caption{The pairwise Pearson correlation heat mat of the numerical features of the $\Delta_w$-optimized n-plets(3, 13, and 20) from Figure 2.a of the main paper, and the pairwise, classical Information Imbalances of the same n-plets.}
    \label{fig:corr_heat}
\end{figure}

The Weighted Information Imbalance approach (main paper) automatically leads to the selection of tuples of features which are practically uncorrelated. This is demonstrated by calculation of the Pearson correlation coefficient and the pairwise classic Information Imbalance (introduced in ref. \cite{glielmo2021ranking}) for all the numerical features contained in the $\Delta_w$-optimized n\-plets (Fig. \ref{fig:corr_heat}.). The mean of pairwise correlations of numerical features in \textit{e.g.} the best 13\-plet (Fig. \ref{fig:corr_heat} blue outline), is $\mean{r} = 0.02$, and the pairwise Information Imbalances mean is $\mean{\Delta} = 0.96$, both of which point towards a high degree of orthogonality. 

\subsection{Information Imbalance between numerical patient features}
\begin{figure}[ht!]
    \centering
    \includegraphics[width=0.8\textwidth]{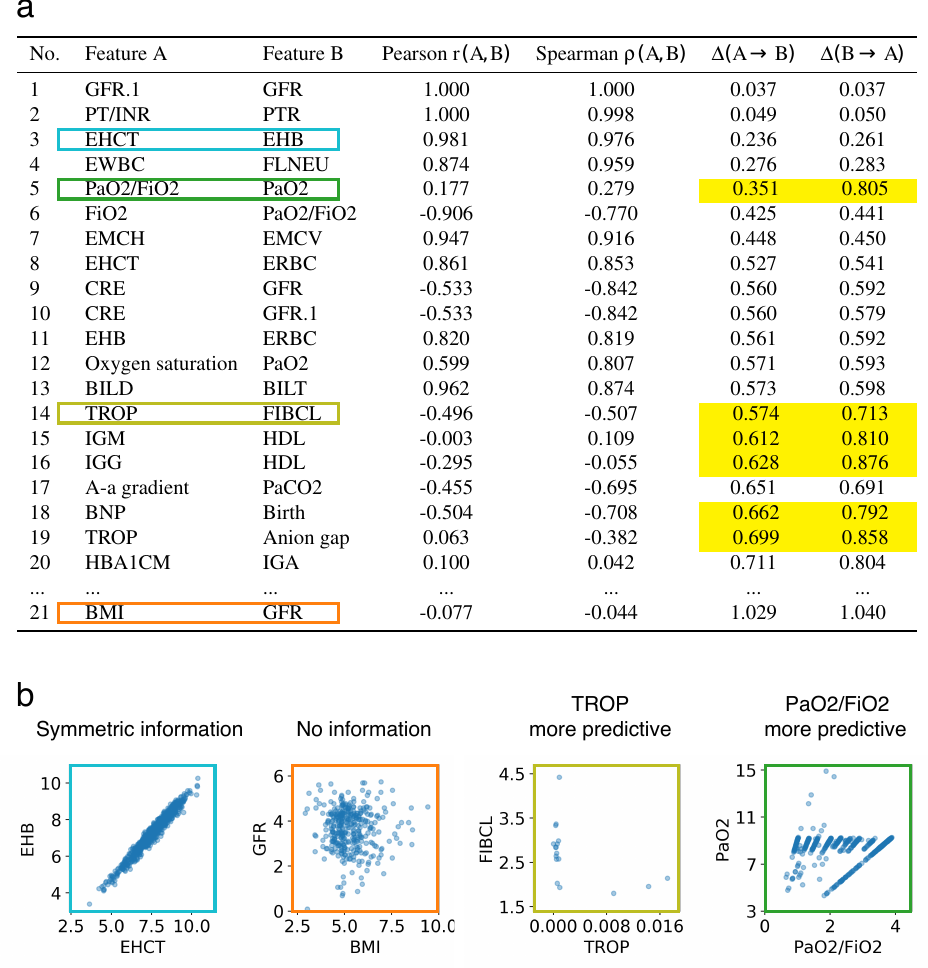}
    \caption{\textbf{a}: Features ordered according to the lowest Information Imbalances towards another feature, and their Pearson and Spearman correlation coefficients. Yellow colored rows have notably asymmetric Information Imbalances, where $|\Delta(A\rightarrow B) -  \Delta(B\rightarrow A)| > 0.1$. \textbf{b}: Scatter plots of several of the feature \textit{vs.} each other from A. The values are the normalized features.}
    \label{fig:results0}
\end{figure}

We also use the classic Information Imbalance to investigate the relationships between the input features. We consider the 90 numerical input features for which it is possible to estimate the standard Information Imbalance introduced in ref. \cite{glielmo2021ranking}. We computed the Information Imbalance $\Delta$ between each pair of features using the implementation in the Python package DADApy \cite{Dada2022}.
$\Delta(A \rightarrow B)$  is close to zero if  feature A predicts feature B  well. It is close to one if feature A does not provide information on feature B.
For each pair of features we also computed  the standard Pearson and Spearman correlation coefficients $r$ and $\rho$, which are $\pm$1 in the case of a perfect positive or negative correlation, and 0 if there is no correlation. If two features correlate strongly, $\Delta(A \rightarrow B)$ and $\Delta(B \rightarrow A)$ should both be small and similar numbers, if both predict each other to an equal amount. However, if one feature predicts  the other, but not vice versa, there exists an asymmetric correlation, and this is reflected in an asymmetric Information Imbalance. This phenomenon, as we will see, is not captured by Pearson and Spearman correlations.
figures/
In the table in Fig. \ref{fig:results0}a, we report the Information Imbalance and the correlation coefficients between the 20 pairs of features with the lowest 
$\Delta(A \rightarrow B)$. To highlight some possible relationships, we plot some of these feature as a function of each other in the bottom panels (Fig. \ref{fig:results0}b).

The Information Imbalance faithfully captures features which have strong correlations with each other. The top-eight positive correlation couples (all $r > 0.8$) are contained in the top-20 Information Imbalances. A clear sanity check is displayed in the first rows: the two different laboratory methods for the glomerular filtration rate (GFR and GFR.1), which hold the same values, display perfect correlations and extremely low Information Imbalances. This is also true for the prothrombin time (- and international normalized) ratios (PT/INR and PTR), where one is just a normalized version of the other. Correlation and Information Imbalance pick up on the linear relationship between hematocrit (EHCT) and hemoglobin (EHB). EHCT is the percentage of volume occupied by red blood cells relative to whole blood, and therefore is often related to hemoglobin (EHB). Also, the strongest negative correlation pairing is in the top-20 imbalance table (row 6). Thus the strongest correlations account for nine rows in the top-20 Information Imbalances. The other eleven rows are made up by pairings which have less strong correlations, but six of them have high asymmetries in their Information Imbalances towards each other (yellow rows in \ref{fig:results0}a), describing a relationship where one variable is more informative about the other than \textit{vice versa}. These six pairings have predominantly very low correlations, showcasing that correlation fails to identify these asymmetric relationships. The effect is especially pronounced in row five, where PaO2/FiO2 (oxygen partial pressure over fractional inspired oxygen) has a low Information Imbalance towards PaO2 (oxygen partial pressure) and such explains this feature space well, while the same is not true \textit{v.v.} This can be used as a proof of concept because indeed PaO2/FiO2 is the value of PaO2 divided by FiO2 (fractional inspired oxygen) - a simple relationship via one confounding variable which is not detected by correlation (r=0.177). It should be noted that, from a clinical point of view, measuring the PaO2/FiO2 ratio can become very challenging: if patients are not on invasive mechanical ventilation, it is almost impossible to know the exact FiO2, because the devices deliver a variable inspired oxygen concentration. Information Imbalance also detected similar cases where the exact relationship is not known: Here we report asymmetric relationships between troponin (TROP), a well-known marker of cardiac injury, and tissue damage marker fibrinogen (FIBCL), as well as between the immunoglobulins IGM / IGG and the high-density lipoprotein (HDL). TROP values are more predictive of FIBCL values than the other way around. Fibrinogen is a plasma acute-phase reactant protein produced by the liver and is a major coagulation factor. Its concentration increases with inflammation, and it is traditionally considered a risk factor for cardiovascular disease \cite{Stulnig_2013, Lowe2004}, which might explain the connection to troponin. Troponin, on the other hand, is a very specific marker: recent studies showed that troponin dosage should be considered as a prognostic indicator in all patients with moderate/severe COVID-19 at hospital admission and in the case of clinical deterioration \cite{Lippi2020}. Retrospective data have placed a strong emphasis on the possibility that acute myocardial injury represents a critical component in the development of serious complications in patients hospitalized with COVID-19 \cite{Chen2020, Shi2020, Guo2020}. To the best of our knowledge, there is no literature concerning the exact relationships between IGM / IGG and HDL.

While causality-based feature selection methods only aim at finding causalities between the predicting features and the target classes \cite{Yu2020}, we point out here that inter-feature asymmetric relationships, such as inter-feature causalities, could also be important. They are not captured by standard correlation analyses, and they could lead to identify redundancy effects when tuples of features are used for predictive purposes. 
This pairwise Information Imbalance between the features was used as a sanity check of the prediction tuples used in the previous section. All $\Delta_w$-optimized tuples were indeed non-redundant, with Information Imbalances of $> 0.85$ between each other.

\section{The stability of selected features the with small random numbers in the input}
\begin{figure*}[ht!]
    \centering
    \includegraphics[width=0.99\textwidth]{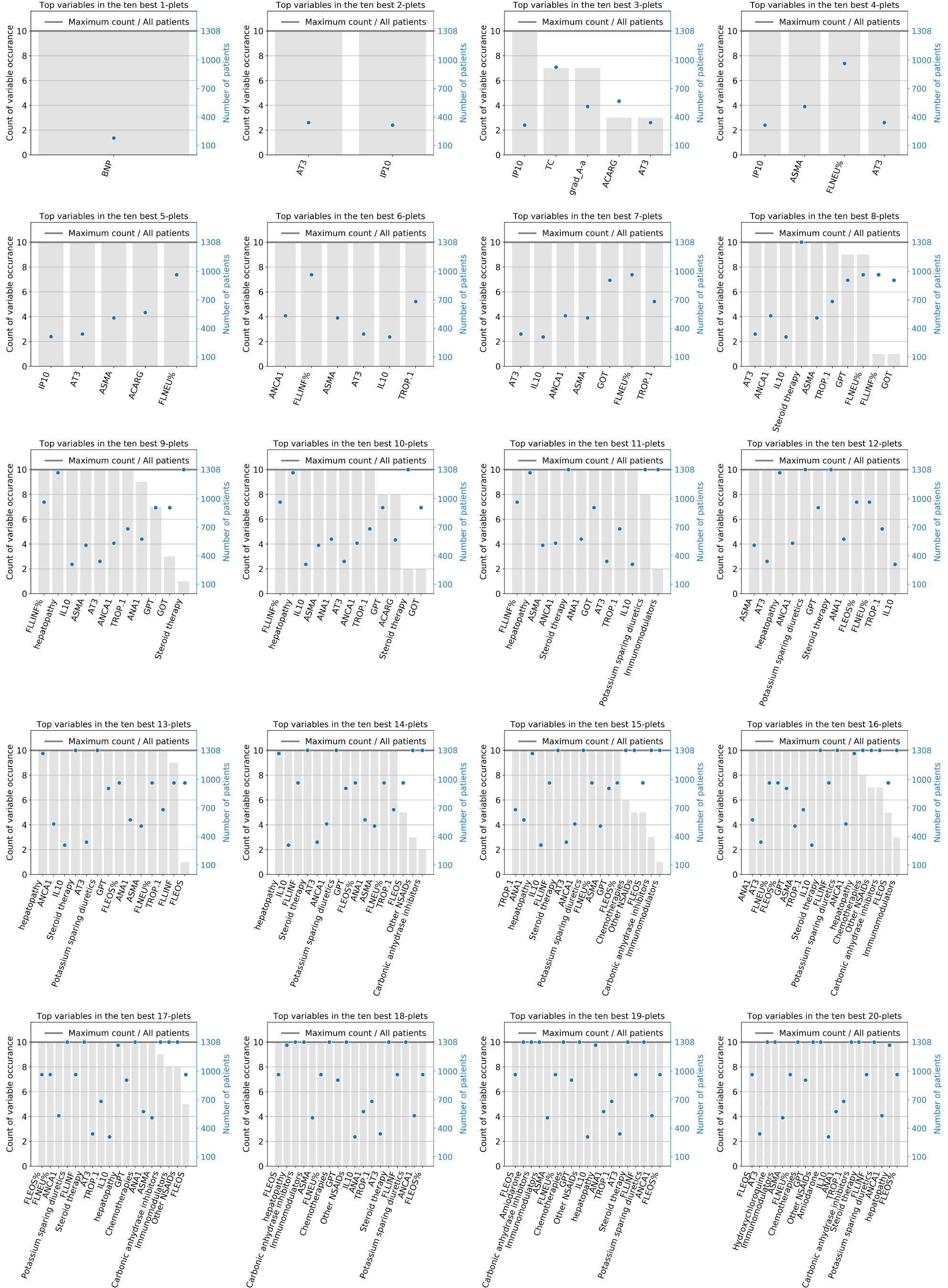}
    \caption{The panels show the variables chosen in each of the $\Delta_w$ feature tuples of different size across ten implementations with different small random numbers added to the degenerate input features. The grey bars corresponding to the left axis shows in how many of the 10 copies a feature was chosen, the blue dots and right axis indicate the number of patients for whom the feature was available.}
    \label{fig:s_10copies}
\end{figure*}

Since the addition of the small random numbers to degenerate inputs renders $\Delta_w$ a stochastic variable, we probed the influence of changing the random seed in ten instances. The Information Imbalances of optimal tuples with the same tuple sizes are mostly identical, up to the second digit, with standard deviations on the order of $10^{-4}$. Also the optimal tuples themselves showed little variability, as seen in Fig. \ref{fig:s_10copies}.

\printbibliography